\documentclass{PoS}

\title{Branching ratios of the pseudoscalar glueball with a mass of 2.6 GeV}

\ShortTitle{Branching ratios of the pseudoscalar glueball with a mass of 2.6 GeV}

\author{\speaker{Walaa I. Eshraim}\thanks{In collaboration with Francesco Giacosa and Dirk H. Rischke}\\
        {Institute for Theoretical Physics, Goethe-University, Max-von-Laue-Str. 1, 
\\60438 Frankfurt am Main, Germany}\\
        E-mail: \email{weshraim@th.physik.uni-frankfurt.de}}
        
\author{{Stanislaus Janowski}\thanks{In collaboration with Francesco Giacosa and Dirk H. Rischke}\\
        {Institute for Theoretical Physics, Goethe-University, Max-von-Laue-Str. 1, 
\\60438 Frankfurt am Main, Germany}\\
        E-mail: \email{janowski@th.physik.uni-frankfurt.de}}

\abstract{We consider a $N_f=3$ chiral Lagrangian which describes the interaction between
the pseudoscalar glueball, $J^{PC}=0^{-+}$, and scalar and pseudoscalar mesons.
We calculate the decay widths of the pseudoscalar glueball, where we fixed its mass to 2.6 GeV,
as predicted by lattice QCD simulations, and take a closer look in the scalar-isoscalar
decaying channel. We present our results as branching ratios which are relevant for the
future PANDA experiment at the FAIR facility.}

\FullConference{Xth Quark Confinement and the Hadron Spectrum,\\
		October 8-12, 2012\\
		TUM Campus Garching, Munich, Germany}

\begin{document}

\section{Introduction}

Glueballs are hypothetical strongly interacting particles. They are made of 
gluons, the gauge bosons of Quantum Chromodynamics (QCD). The reason for the expectation
of such fundamental objects in the nature is the nonabelian structure of QCD.
In this context the study of glueballs is an important field of research in hadronic physics,
relevant for the understanding of the structure of some experimentally verified
resonances and the phenomenological description of low-energy QCD, see Ref.
\cite{review} and references therein.

Lattice QCD calculations predict a complete glueball spectrum \cite{Morningstar}, 
where the third lightest glueball, which is investigated in this work,
is a pseudoscalar state ($J^{PC}=0^{-+}$) with a mass of about 2.6 GeV. We
study the decays of this pseudoscalar glueball, denoted as $\tilde{G}\equiv gg$,
into three pseudoscalar mesons, $\tilde{G}\rightarrow PPP$,
and a pseudoscalar and a scalar meson, $\tilde{G}\rightarrow PS$.

The effective chiral Lagrangian introduced in Refs. \cite{our,proc,proc2} contains
the relevant tree-level vertices necessary for evaluating the corresponding decay
widths. Due to the assignment issue of the bare scalar-isoscalar states, and their
mixing which generates the resonances $f_{0}(1370)$, $f_{0}(1500)$ and $f_{0}(1710)$, 
e.g. Refs. \cite{scalars,G12,stani,dick}, we consider the 
decays of the pseudoscalar glueball $\tilde{G}$ into $\eta$ and $\eta^{\prime}$, 
respectively and a scalar-isoscalar in more detail. 
Our numerical results are given as branching ratios in order to present
a parameter free prediction of our approach.

The PANDA experiment at the FAIR facility in Darmstadt is currently under
construction and will use an 1.5 GeV antiproton beam hitting a proton target
at rest \cite{panda}. Therefore a centre of mass energy higher than $\sim$2.5
GeV will be reached and the 2.6 GeV pseudoscalar glueball can be directly
produced as an intermediate state \cite{proc}.

\section{The chiral Lagrangian}

The interaction between the pseudoscalar glueball $\tilde{G}\equiv gg$ with
the quantum numbers $J^{PC}=0^{-+}$ and the ordinary scalar and pseudoscalar
mesons is described by the Lagrangian \cite{our,proc,schechter}:%

\begin{equation}
\mathcal{L}_{\tilde{G}}^{int}=ic_{\tilde{G}\Phi}\tilde{G}\left(
{\textrm{det}}\Phi-{\textrm{det}}\Phi^{\dag}\right)\label{intlag},
\end{equation}
where $c_{\tilde{G}\Phi}$ is a coupling constant and $\Phi$ is a multiplet containing the ordinary scalar
and pseudoscalar mesons. In this study we consider three flavours, $N_{f}=3$, thus $c_{\tilde{G}\Phi}$ is dimensionless
and the multiplet $\Phi$ reads \cite{dick}:

\begin{equation}
\Phi=\frac{1}{\sqrt{2}}\left(
\begin{array}
[c]{ccc}%
\frac{(\sigma_{N}+a_{0}^{0})+i(\eta_{N}+\pi^{0})}{\sqrt{2}} & a_{0}^{+}%
+i\pi^{+} & K_{S}^{+}+iK^{+}\\
a_{0}^{-}+i\pi^{-} & \frac{(\sigma_{N}-a_{0}^{0})+i(\eta_{N}-\pi^{0})}%
{\sqrt{2}} & K_{S}^{0}+iK^{0}\\
K_{S}^{-}+iK^{-} & \bar{K}_{S}^{0}+i\bar{K}^{0} & \sigma_{S}+i\eta_{S}%
\end{array}
\right). \label{phimatex}%
\end{equation}

As discussed in Refs. \cite{our,proc,proc2}, let us briefly consider the
symmetries of the effective Lagrangian (\ref{intlag}). The pseudoscalar
glueball $\tilde{G}$ is made of gluons and is therefore chirally invariant.
The multiplet $\Phi$ transforms under the chiral symmetry as $\Phi\rightarrow
U_{L}\Phi U_{R}^{\dagger}$, where $U_{L(R)}=e^{-i\theta_{L(R)}^{a}t^{a}}$ is
an element of $U(3)_{L(R)}$. Performing these transformations on the
determinant of $\Phi$ it is easy to prove that this object is invariant under
$SU(3)_{L}\times SU(3)_{R}$, but not under the axial $U_{A}(1)$
transformation:
\begin{equation}
\mathrm{det}\Phi\rightarrow\mathrm{det}U_{A}\Phi U_{A}=e^{-i\theta_{A}%
^{0}\sqrt{2N_{f}}}\mathrm{det}\Phi\neq\mathrm{det}\Phi\,.\label{ua}%
\end{equation}
This is in agreement with the chiral anomaly. Consequently the effective
Lagrangian (\ref{intlag}) possesses only the $SU(3)_{L}\times SU(3)_{R}$
symmetry. Further essential symmetries of the strong interacting matter are
the parity $\mathcal{P}$ and charge conjugation $\mathcal{C}$. The
pseudoscalar glueball and the multiplet $\Phi$ transform under parity as
\begin{equation}
\tilde{G}(t,\vec{x})\rightarrow-\tilde{G}(t,-\vec{x})\,,\ \Phi(t,\vec
{x})\rightarrow\Phi^{\dagger}(t,-\vec{x})\,,\label{p}%
\end{equation}
and under charge conjugation as
\begin{equation}
\tilde{G}\rightarrow\tilde{G}\,,\ \Phi\rightarrow\Phi^{T}\,.\label{c}%
\end{equation}
Performing these discrete transformations $\mathcal{P}$ and $\mathcal{C}$ on
the effective Lagrangian (\ref{intlag}) leave it unchanged. In conclusion, one
can say that the symmetries of the effective Lagrangian (\ref{intlag}) are in
agreement with the symmetries of the QCD Lagrangian.

Let us now consider the assignment of the ordinary mesonic d.o.f. in Eq.
(\ref{intlag}) or (\ref{phimatex}). In the pseudoscalar sector we
assign the fields $\vec{\pi}$ and $K$ to the physical pion isotriplet and the
kaon isodoublet \cite{PDG}. The bare quark-antiquark fields $\eta_{N}%
\equiv(\bar{u}u+\bar{d}d)/\sqrt{2}$ and $\eta_{S}\equiv\ \bar{s}s$ are the
nonstrange and strange mixing contributions of the physical states $\eta$ and
$\eta^{\prime}$ \cite{dick}. In the effective Lagrangian (\ref{intlag}) there
exist a mixing between the bare pseudoscalar glueball $\tilde{G}$ and the both
bare fields $\eta_{N}$ and $\eta_{S}$, but, due to the large mass difference
between the pseudoscalar glueball and the pseudoscalar quark-antiquark fields,
it turns out that its mixing is very small and is therefore negligible.
In the scalar sector the field $\vec{a}_{0}$ corresponds to the physical
isotriplet state $a_{0}(1450)$ and the scalar kaon field $K_{S}$ to the
physical isodoublet state $K_{0}^{\star}(1430)$ \cite{PDG}. 
The field $\sigma_{N}\equiv(\bar{u}u+\bar{d}d)/\sqrt{2}$
is the bare nonstrange isoscalar field and it corresponds to the resonance
$f_{0}(1370)$ \cite{dick,Cheng}. The field $\sigma_{S}\equiv\bar{s}s$ is the
bare strange isoscalar field and the debate about its assignment to a physical
state is still ongoing; in a first approximation it can be assigned to the resonance
$f_{0}(1710)$ \cite{dick} or $f_{0}(1500)$ \cite{Cheng}. (Scalars
below 1 GeV are predominantly tetraquarks or mesonic molecular states, see
Refs. \cite{tetraquark,lowscalars} and references therein, and are not
considered here). In order to properly 
take into account mixing effects in the scalar-isoscalar sector, we have also used
the results of Refs. \cite{G12,Cheng}. The mixing takes the form:

\begin{equation}
\left(
\begin{array}
[c]{c}%
\,f_{0}(1370)\\
\,f_{0}(1500)\\
\,f_{0}(1710)
\end{array}
\right)  =B\cdot\left(
\begin{array}
[c]{c}%
\sigma_{N}\equiv \bar{n}n =(\bar{u}u+\bar{d}d)/\sqrt{2}\\
G\equiv gg\\
\sigma_{S}\equiv\bar{s}s
\end{array}
\right)  \label{isomix} ,%
\end{equation}%
where $B$ is an orthogonal (3 $\times$ 3) matrix and $G\equiv gg$ a scalar
glueball field which is absent in this study.

In accordance with the spontaneous breaking of the chiral symmetry we shift
the scalar-isoscalar fields by their vacuum expectation values $\sigma
_{N}\rightarrow\sigma_{N}+\phi_{N}$ and $\sigma_{S}\rightarrow\sigma_{S}%
+\phi_{S}$, where $\phi_{N}$ and $\phi_{S}$ are the corresponding chiral
condensates. In order to be consistent with the full effective chiral
Lagrangian of the extended Linear Sigma Model \cite{dick,Susanna,Paper1} we
have to consider the shift of the axial-vector fields and thus to redefine
the wave function of the pseudoscalar fields
\begin{equation}
\vec{\pi}\rightarrow Z_{\pi}\vec{\pi}\,,\ K\rightarrow Z_{K}K\,,\ \eta
_{N,S}\rightarrow Z_{\eta_{N,S}}\eta_{N,S}\;,\label{psz}%
\end{equation}
where $Z_{i}$ are the renormalization constants of the corresponding wave
functions \cite{dick}.

\section{Results}

From the effective chiral Lagrangian (\ref{intlag}) with the unknown coupling
constant $c_{\tilde{G}\Phi}$ we calculated the two- and three-body decays of
the pseudoscalar glueball $\tilde{G}$. We present the decay widths as
branching ratios in order to obtain parameter free results as predictions
for the future PANDA experiment at the FAIR facility. For our
calculations we fixed the mass of the pseudoscalar glueball to
$M_{\tilde{G}}=2.6$ GeV. This value is obtained by studying of the pure Yang-Mills sector
on lattice QCD \cite{Morningstar}. Due to the mixing in the scalar-isoscalar
channel we evaluated explicitly the decays of the pseudoscalar glueball
$\tilde{G}$ into $\eta$ and $\eta^{\prime}$ and one of the
scalar-isoscalar resonances $f_{0}(1370),f_{0}(1500)$ or $f_{0}(1710)$. We
used for the transformation matrix $B$ in Eq. (\ref{isomix}) the solution (1)
and (2) of Ref. \cite{G12}:

\begin{equation}
B_{1}=
\left(
\begin{array}
[c]{ccc}%
0.86 & 0.45 & 0.24\\
-0.45 & 0.89 & -0.06\\
-0.24 & -0.06 & 0.97
\end{array}
\right)  \label{s1} ,
\end{equation}

\begin{equation}
B_{2}=
\left(
\begin{array}
[c]{ccc}%
0.81 & 0.54 & 0.19\\
-0.49 & 0.49 & 0.72\\
0.30 & -0.68 & 0.67
\end{array}
\right)  \label{s2}
\end{equation}
and the solution of Ref. \cite{Cheng}:%

\begin{equation}
B_{3}=\left(
\begin{array}
[c]{ccc}%
0.78 & -0.36 & 0.51\\
-0.54 & 0.03 & 0.84\\
0.32 & 0.93 & 0.18
\end{array}
\right)  \label{s3} .%
\end{equation}%

In the solution 1 of Ref. \cite{G12} the resonance $f_{0}(1370)$ is
predominantly $\bar{n}n$ state, the resonance $f_{0}(1500)$ is
predominantly a glueball, and $f_{0}(1710)$ is predominantly a strange $\bar
{s}s$ state. In the solution 2 of Ref. \cite{G12} and in the solution of Ref.
\cite{Cheng} the resonance $f_{0}(1370)$ is still predominantly nonstrange
$\bar{n}n$ state, but $f_{0}(1710)$ is now predominantly a glueball, and
$f_{0}(1500)$ predominantly a strange $\bar{s}s$ state.

In Table 1 we present our results for the decay channels $\tilde{G}\rightarrow
PPP$ and in Table 2 and 3 we show the results for the decay channels
$\tilde{G}\rightarrow PS$, whereby in Table 3 the bare scalar-isoscalar
states are substituted by the physical ones \cite{our,proc,proc2}.
$\Gamma_{\tilde{G}}^{tot}=\Gamma_{\tilde{G}\rightarrow PPP}+\Gamma_{\tilde
{G}\rightarrow PS}$ is the total decay width. In Table 2 the
decays $\tilde{G}\rightarrow\eta\sigma_{S}$ and $\tilde
{G}\rightarrow\eta^{\prime}\sigma_{S}$ correspond to the assignment
$\sigma_{S}\equiv f_{0}(1710)$ and to $\sigma_{S}\equiv f_{0}(1500)$
(values in the brackets). Finally, in Table 3 we present results
in the scalar-isoscalar sector where the mixing matrices (\ref{s1}), (\ref{s2}), 
and (\ref{s3}) are taken into account.

The largest contribution to the total decay width is given by the following
decay channels: $KK\pi$, which contributes to almost 50\%, as well as $\eta
\pi\pi$ and $\eta^{\prime}\pi\pi$, where each one contributes of about 10\%.
The substitution of the bare scalar-isoscalar states through the resonances
$f_0(1370)$, $f_0(1500)$ and $f_0(1710)$ did not affect the several decay
channels. The contribution of the pseudoscalar glueball decay into 
the scalars-isoscalars is still about 5\% independently on the mixing scenario.
A further interesting outcome of our approach is that the decay of
the pseudoscalar glueball into three pions, $\tilde{G}\rightarrow\pi\pi\pi$,
is not allowed.

\begin{center}%
\begin{table}[h] \centering
\begin{tabular}
[c]{|c|c|}\hline
Quantity & Value\\\hline
$\Gamma_{\tilde{G}\rightarrow KK\eta}/\Gamma_{\tilde{G}}^{tot}$ &
$0.049$\\\hline
$\Gamma_{\tilde{G}\rightarrow KK\eta^{\prime}}/\Gamma_{\tilde{G}}^{tot}$ &
$0.019$\\\hline
$\Gamma_{\tilde{G}\rightarrow\eta\eta\eta}/\Gamma_{\tilde{G}}^{tot}$ &
$0.016$\\\hline
$\Gamma_{\tilde{G}\rightarrow\eta\eta\eta^{\prime}}/\Gamma_{\tilde{G}}^{tot}$
& $0.0017$\\\hline
$\Gamma_{\tilde{G}\rightarrow\eta\eta^{\prime}\eta^{\prime}}/\Gamma_{\tilde
{G}}^{tot}$ & $0.00013$\\\hline
$\Gamma_{\tilde{G}\rightarrow KK\pi}/\Gamma_{\tilde{G}}^{tot}$ &
$0.47$\\\hline
$\Gamma_{\tilde{G}\rightarrow\eta\pi\pi}/\Gamma_{\tilde{G}}^{tot}$ &
$0.16$\\\hline
$\Gamma_{\tilde{G}\rightarrow\eta^{\prime}\pi\pi}/\Gamma_{\tilde{G}}^{tot}$ &
$0.095$\\\hline
\end{tabular}%
\caption
{Branching ratios for the decay of the pseudoscalar glueball with the mass of $M_{\tilde
{G}}=2.6$ GeV into three pseudoscalar mesons: $\tilde{G} \rightarrow PPP$.}%
\end{table}%
\end{center}

\begin{table}[h] \centering
\begin{tabular}
[c]{|c|c|}\hline
Quantity & Value\\\hline
$\Gamma_{\tilde{G}\rightarrow KK_{S}}/\Gamma_{\tilde{G}}^{tot}$ &
$0.060$\\\hline
$\Gamma_{\tilde{G}\rightarrow\pi a_{0}}/\Gamma_{\tilde{G}}^{tot}$ &
$0.083$\\\hline
$\Gamma_{\tilde{G}\rightarrow\eta\sigma_{N}}/\Gamma_{\tilde{G}}^{tot}$ &
$0.0000026$\\\hline
$\Gamma_{\tilde{G}\rightarrow\eta^{\prime}\sigma_{N}}/\Gamma_{\tilde{G}}%
^{tot}$ & $0.039$\\\hline
$\Gamma_{\tilde{G}\rightarrow\eta\sigma_{S}}/\Gamma_{\tilde{G}}^{tot}$ &
$0.012$ $(0.015)$\\\hline
$\Gamma_{\tilde{G}\rightarrow\eta^{\prime}\sigma_{S}}/\Gamma_{\tilde{G}}%
^{tot}$ & $0$ $(0.0082)$\\\hline
\end{tabular}%
\caption
{Branching ratios for the decay of the pseudoscalar glueball with the mass of $M_{\tilde
{G}}=2.6$ GeV into a pseudoscalar and a scalar meson: $\tilde{G} \rightarrow
PS$. In the last two rows $\sigma_S$ is assigned to $ f_0(1710)$ 
or to $ f_0(1500)$ (values in the brackets) }%
\end{table}%
%

\begin{table}[h] \centering
\begin{tabular}
[c]{|c|c|c|c|}\hline
Quantity & Sol. 1 of Ref \cite{G12} & Sol. 2 of Ref \cite{G12} & Sol. of Ref
\cite{Cheng}\\\hline
$\Gamma_{\tilde{G}\rightarrow\eta f_{0}(1370)}/\Gamma_{\tilde{G}}^{tot}$ &
\multicolumn{1}{|l|}{$0.00093$} & \multicolumn{1}{|l|}{$0.00058$} &
\multicolumn{1}{|l|}{$0.0044$}\\\hline
$\Gamma_{\tilde{G}\rightarrow\eta f_{0}(1500)}/\Gamma_{\tilde{G}}^{tot}$ &
\multicolumn{1}{|l|}{$0.000046$} & \multicolumn{1}{|l|}{$0.0082$} &
\multicolumn{1}{|l|}{$0.011$}\\\hline
$\Gamma_{\tilde{G}\rightarrow\eta f_{0}(1710)}/\Gamma_{\tilde{G}}^{tot}$ &
\multicolumn{1}{|l|}{$0.011$} & \multicolumn{1}{|l|}{$0.0053$} &
\multicolumn{1}{|l|}{$0.00037$}\\\hline
$\Gamma_{\tilde{G}\rightarrow\eta^{\prime}f_{0}(1370)}/\Gamma_{\tilde{G}%
}^{tot}$ & \multicolumn{1}{|l|}{$0.038$} & \multicolumn{1}{|l|}{$0.033$} &
\multicolumn{1}{|l|}{$0.043$}\\\hline
$\Gamma_{\tilde{G}\rightarrow\eta^{\prime}f_{0}(1500)}/\Gamma_{\tilde{G}%
}^{tot}$ & \multicolumn{1}{|l|}{$0.0062$} & \multicolumn{1}{|l|}{$0.00020$} &
\multicolumn{1}{|l|}{$0.00013$}\\\hline
$\Gamma_{\tilde{G}\rightarrow\eta^{\prime}f_{0}(1710)}/\Gamma_{\tilde{G}%
}^{tot}$ & \multicolumn{1}{|l|}{$0$} & \multicolumn{1}{|l|}{$0$} &
\multicolumn{1}{|l|}{$0$}\\\hline
\end{tabular}%
\caption{Branching ratios for the decays of the pseudoscalar glueball $\tilde
{G}$ into $\eta$ and $\eta
'$, respectively and one of the scalar-isoscalar states: $f_0(1370), f_0(1500)$ and $ f_0(1710)$
by using three different mixing scenarios of these scalar-isoscalar states \cite{G12,Cheng}.
The mass of the pseudoscalar glueball is $M_{\tilde{G}}=2.6$ GeV.}%
\end{table}%

\section{Conclusions}

We have presented a chirally invariant three-flavour effective Lagrangian with
scalar and pseudoscalar quark-antiquark states and a pseudoscalar glueball. We
have calculated two- and three-body decay processes of the pseudoscalar
glueball with a mass of 2.6 GeV, as evaluated by lattice QCD. It turns out 
that our results depend only slightly on the scalar-isoscalar mixing.
We predict that the decay channel $\tilde{G}\rightarrow KK\pi$ is the largest one and $\tilde
{G}\rightarrow\eta\pi\pi$ as well as $\tilde{G}\rightarrow\eta^{\prime}\pi\pi$
are the next dominant ones \cite{our,proc}. Moreover, the decay channel
$\tilde{G}\rightarrow\pi\pi\pi$ is not allowed. \newline The results presented
in this work can be tested in the upcoming PANDA experiment at the FAIR
facility \cite{panda}.

\section*{Acknowledgments}

The authors thank Francesco Giacosa and Dirk H. Rischke for useful
discussions. W.E.\ acknowledges support from DAAD and HGS-HIRe,
S.J.\ acknowledges support from H-QM and HGS-HIRe.

\end{document}